# Buyer Artificial Intelligence-Enabled Environmental Governance and Supplier Environmental Controversies: An Organizational Information Processing and Signaling Perspective

**Yongchao Martin Ma, *Senior Member, IEEE*** [*]
School of Management
Huazhong University of Science & Technology, PR China
Department of Marketing
City University of Hong Kong, Hong Kong

**Xinya Guan** [†]
School of Economics and Management
Harbin Engineering University
Harbin, PR China

## Abstract

Environmental controversies in global supply chains pose significant risks for global buyers. This study examines whether overseas suppliers' exposure to buyers' artificial intelligence (AI)-enabled environmental governance reduces supplier environmental controversies. Drawing on organizational information processing theory and signaling theory, we investigate how suppliers' exposure to AI-enabled governance influences their environmental controversies and the institutional contingencies under which this effect varies. Using text analysis to measure buyer AI-enabled environmental governance, we analyze panel data on 2,505 suppliers of U.S.-listed firms across 41 countries from 2020 to 2024 with multidimensional fixed-effects models. We find that suppliers' exposure to buyer AI-enabled environmental governance is negatively associated with supplier environmental controversies in the following year. This negative relationship is stronger in supplier countries with higher AI readiness and regulatory quality. The study contributes to research on AI-enabled sustainability governance and sustainable supply chain risk management.

***Keywords*** AI-enabled environmental governance · supplier environmental controversies · sustainable supply chain management · organizational information processing theory · signaling theory · AI readiness

[*]This work was supported by the National Natural Science Foundation of China (No. 72602137), the Hong Kong Scholars Program (No. XJ2026002), and the China Postdoctoral Science Foundation - Hubei Joint Support Program (No. 2025T100HB). The part of the computation is completed in the HPC Platform of Huazhong University of Science and Technology.
[†]Corresponding author. Email: `guanxinya@hrbeu.edu.cn`

# Buyer Artificial Intelligence-Enabled Environmental Governance and Supplier Environmental Controversies: An Organizational Information Processing and Signaling Perspective

## ABSTRACT

Environmental controversies in global supply chains pose significant risks for global buyers. This study examines whether overseas suppliers' exposure to buyers' artificial intelligence (AI)-enabled environmental governance reduces supplier environmental controversies. Drawing on organizational information processing theory and signaling theory, we investigate how suppliers' exposure to AI-enabled governance influences their environmental controversies and the institutional contingencies under which this effect varies. Using text analysis to measure buyer AI-enabled environmental governance, we analyze panel data on 2,505 suppliers of U.S.-listed firms across 41 countries from 2020 to 2024 with multidimensional fixed-effects models. We find that suppliers' exposure to buyer AI-enabled environmental governance is negatively associated with supplier environmental controversies in the following year. This negative relationship is stronger in supplier countries with higher AI readiness and regulatory quality. The study contributes to research on AI-enabled sustainability governance and sustainable supply chain risk management.

## 1. Introduction

Increasing pressure on global buyers to ensure responsible practices in their supply chains has heightened these firms' exposure to risks stemming from supplier environmental irresponsibility (Damberg et al., 2022; Hajmohammad et al., 2024; Kähkönen et al., 2023; Kim et al., 2015). High-profile cases, such as Nestlé's rainforest destruction through its supplier Sinar Mas, have intensified stakeholder pressure on buyers to address environmental incidents across their global supply chains. However, traditional supplier governance mechanisms often struggle to provide timely and reliable oversight, as supplier operations span multiple jurisdictions characterized by fragmented environmental information, limited visibility, and heterogeneous institutional environments (Fahimnia et al., 2015; Kach et al., 2025; Testa & Iraldo, 2010). Against this backdrop, artificial intelligence (AI) technologies have created new possibilities for scalable environmental governance by enabling firms to gather, interpret, and synthesize complex information across organizational and national boundaries (Jackson et al., 2024; Li et al., 2026; Mishra et al., 2024). Prior research has documented the implications of AI for corporate sustainability and environmental, social, and governance (ESG) performance (Bai et al., 2020; Dimes et al., 2026; Fosso Wamba et al., 2024; Podrecca et al., 2024). However, little is known about whether AI-enabled governance extends beyond focal firms to influence supplier environmental behavior.

AI technologies are increasingly being incorporated into sustainability and ESG practices (Singh et al., 2024; Wu, 2026; Xiao & Xiao, 2025). AI techniques, such as machine learning, natural language processing, computer vision, and Generative AI, can learn from vast quantities of high-dimensional data (including text, speech, and image data) and perform high-skilled, non-routine tasks, such as prediction, detection, and classification (Babina et al., 2024; Bhattacharya et al., 2024; Jackson et al., 2024; Qian et al., 2026). Collectively, these capabilities enable firms to collect, integrate, and process fragmented environmental information in real time. For example, Schneider Electric, one of the world's most sustainable companies, has invested in agentic AI–AI systems that actively adapt and optimize for emissions reduction, predictive maintenance, and resource efficiency. Although rooted in buyers' internal environmental governance systems, AI-enabled environmental governance may extend beyond firm boundaries to suppliers in dispersed supply networks by enhancing environmental information processing and signaling greater oversight (Rehman et al., 2026; Vann Yaroson et al., 2026). Prior studies have primarily focused on AI's implications for focal firms' ESG outcomes (Rehman et al., 2026; Tian et al., 2025). Although prior studies suggest that macro-level AI policies may generate environmental spillovers across supply-chain relationships (Yin et al., 2026), limited attention has been paid to whether AI-enabled governance can influence environmental conduct beyond firm boundaries.

Furthermore, the cross-border spillover effects of AI-enabled environmental governance are likely to depend on institutional environments, which serve as primary drivers of environmental management (Lee et al., 2024; Podrecca & Culot, 2026; Testa et al., 2018). Focusing on foreign suppliers as the empirical context, this study seeks to examine whether buyer firms' AI-enabled environmental governance influences supplier environmental controversies and how institutional environments condition this relationship.

To answer this question, we draw on organizational information processing theory (OIPT) and signaling theory, as the spillover effects of AI-enabled environmental governance on supplier environmental controversies may depend on both buyers' information-processing capacity and suppliers' responses to governance signals. We conceptualize buyers' AI-enabled environmental governance as the use of firm-wide AI techniques to gather, interpret, and synthesize environmental information. From an organizational information processing perspective, AI techniques may strengthen buyers' ability to process complex environmental information across organizational and national boundaries. From a signaling perspective, buyers' disclosure of AI-enabled sustainability governance may signal stronger oversight and monitoring capabilities to suppliers, potentially increasing perceived governance pressures.

To provide further support for this theoretical logic, we examine two institutional contingencies that may strengthen buyers' information-processing capacity and suppliers' perceived governance pressures associated with buyer AI-enabled environmental governance. Supplier-country AI readiness, referring to the country-level technical capacity to integrate AI into the delivery of public services and to support broader economic transformation, would enhance buyers' AI-enabled information processing capacities by providing more advanced digital and data infrastructures. Supplier-country regulatory quality, referring to the overall effectiveness of the institutional and regulatory environment in suppliers' home countries, would increase perceived governance pressures by enhancing the credibility and enforceability of environmental oversight while also facilitating more effective environmental information processing.

Taking supplier-year as the unit of analysis, we constructed an unbalanced panel dataset of 2,505 foreign suppliers of non-financial, U.S.-listed firms by merging multiple data sources across 41 countries, including SEC 10-K filings, FactSet Supply Chain Relationships, RepRisk, Oxford Insights, Worldwide Governance Indicators (WGI), and other firm- and country-level databases. We measured buyer AI-enabled environmental governance through sentence-level textual analysis of buyers' 10-K filings and aggregated the resulting measure to the supplier level. We conducted a multi-dimensional fixed-effect analysis with firm-level clustered standard errors to examine our hypotheses and validated our results with a comprehensive set of robustness checks, including alternative measures of the primary variables, alternative samples, alternative estimation approaches, and methods to mitigate endogeneity concerns. Our findings reveal that suppliers' exposure to buyer AI-enabled environmental governance is negatively associated with suppliers' environmental controversies the following year, an association enhanced by higher supplier-country AI readiness and regulatory quality.

Our study contributes to the literature on AI-enabled sustainability governance and sustainable supply chain risk management. First, in contrast to the prior focus on the impacts of AI adoption on firms' ESG outcomes (Dimes et al., 2026; Tian et al., 2025), we draw on OIPT and signaling theory to explore how buyer AI-enabled environmental governance generates cross-border governance spillovers that reduce foreign suppliers' environmental controversies. We further identify important institutional contingencies, namely supplier-country AI readiness and regulatory quality. Second, we contribute to the emerging literature on AI-enabled sustainability governance (Rehman et al., 2026; Xiao & Xiao, 2025) by developing a firm-level measure of AI-enabled environmental governance using sentence-level textual analysis of SEC 10-K filings. This approach provides a scalable way to capture firms' AI-enabled environmental governance practices and offers methodological tools for future research on AI and sustainability. Third, we contribute to the sustainable supply chain risk management literature (Damberg et al., 2022; L. Wang et al., 2025) by shifting attention from traditional supplier governance mechanisms toward AI-enabled environmental governance as a scalable approach to overseeing suppliers' environmental practices across global supply chains. By demonstrating its effectiveness and boundary conditions, we show how AI-enabled governance complements and extends conventional monitoring and supplier oversight approaches.

We also offer implications for managers and policymakers seeking to leverage AI for long-term sustainable development. First, buyer firms should adopt and disclose AI-enabled environmental governance practices to strengthen their capacity to collect, integrate, and monitor environmental information across global supply chains. Such disclosures can also signal heightened environmental oversight and increase suppliers' perceived likelihood of detection for environmental misconduct. Second, policymakers in supplier countries should invest in AI infrastructure and strengthen regulatory quality, as these conditions enhance the effectiveness and credibility of buyers' AI-enabled environmental governance. Finally, our findings highlight the importance of cross-border collaboration between firms and policymakers in improving environmental oversight and sustainable supply chain governance.

The remainder of this paper is organized as follows. Section 2 presents the literature review and develops the hypotheses. Section 3 describes the data and methodology, before Section 4 provides empirical results. Section 5 summarizes the theoretical and managerial implications, as well as limitations and future research, and the last section concludes.

## 2. Literature Review and Hypothesis Development

### 2.1 Sustainable Supply Chain Risks and AI-Enabled Sustainability Governance

At the intersection of sustainable supply chain management and supply chain risk management literatures, supply chain sustainability risk management research emphasizes that stakeholders often attribute responsibility for suppliers' sustainability misconduct to buying firms and may subsequently penalize them (Both & Wilhelm, 2026; Mateska et al., 2025; Sarkis et al., 2010). Media stakeholders, in particular, detect and publicize such misconduct, shaping the perceptions of broader stakeholder audiences (Lin et al., 2025; L. Wang et al., 2025). Prior research has documented the negative consequences of media coverage of supplier sustainability incidents for buying firms, including heightened reputation risk (Kumar et al., 2019), adverse customer reactions (Bregman et al., 2015; Mukandwal et al., 2024; Nichols et al., 2019), stock market penalties (Jacobs & Singhal, 2017; Mateska et al., 2023), and deteriorating financial performance (Lin et al., 2025). Consequently, firms are expected to deploy effective risk management strategies to mitigate downside risks (Hajmohammad et al., 2024; Hajmohammad et al., 2021; Hajmohammad & Vachon, 2016; Mateska et al., 2025). However, traditional sustainable supply chain practices struggle to gather and process information across organizational and national boundaries, a challenge that is particularly salient when environmental information is fragmented, unstructured, and dispersed across supply networks (Kach et al., 2025).

AI techniques concern how computer systems can emulate human intelligence, learn, and adapt to their environment (Xiao & Xiao, 2025), thereby enabling firms to process complex environmental information. Core technologies—such as machine learning, natural language processing, computer vision, and GenAI—can learn from vast quantities of high-dimensional data (including text, speech, and image data), assisting companies in performing high-skilled, non-routine tasks such as prediction, detection, and classification (Agrawal et al., 2019). These capacities position AI as a driver of firm growth and product innovation as well as supply chain management (Herold et al., 2025; Koh et al., 2019; Spreitzenbarth et al., 2024; Y. Wang et al., 2025) , and render its implications for sustainability promising. For example, machine learning techniques can assist companies in identifying potential environmental risks within large datasets and providing intelligent solutions, while natural language processing technologies facilitate real-time monitoring and analysis of societal sentiment in media. Additionally, GenAI can be used to create and assess sustainability disclosures (Dimes et al., 2026).

Previous research indicates that macro-country-level AI readiness facilitates managing carbon transition risk (Vann Yaroson et al., 2026). At the firm level, corporate AI adoption can enable firms to foster long-term sustainable growth by optimizing resource use, operationalizing sustainability-oriented knowledge, improving information transparency, and fostering corporate innovation (Rehman et al., 2026; Tian et al., 2025; Xiao & Xiao, 2025). For example, AI-driven analytics can identify inefficiencies in energy consumption and improve firms' environmental practices. Despite providing valuable insights into AI for sustainability, prior work focuses on the impacts of AI adoption on focal firms' ESG performance, overlooking that AI-enabled sustainability governance can spill over beyond organizational and national boundaries.

This study conceptualizes buyer AI-enabled environmental governance as a firm-level capability through which buyers use AI-enabled systems to collect, integrate, interpret, and synthesize environmental information within and across organizational boundaries. Although this capability originates within buyer firms, it may generate governance spillovers across dispersed supply networks by enhancing environmental visibility and oversight beyond organizational boundaries. To explain how buyer AI-enabled environmental governance generates cross-border governance spillovers, we draw on organizational information processing and signaling theory to develop our hypotheses, which we elaborate in the next section. The conceptual model is summarized in Figure 1.

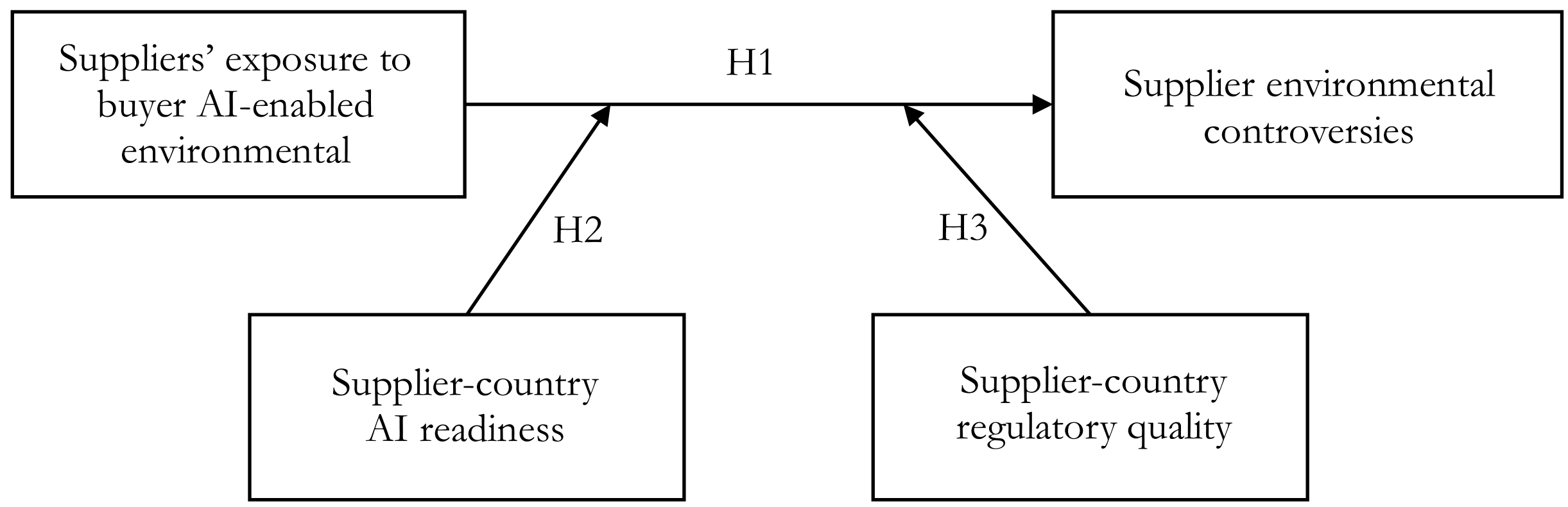

**Figure 1.** Conceptual model

**2.2 Buyer AI-Enabled Environmental Governance and Supplier Environmental Controversies**

Suppliers are often embedded within broader buyer networks and maintain relationships with multiple buyers simultaneously (Foerstl et al., 2015), and may therefore be exposed to varying levels of buyer AI-enabled environmental governance. Such exposure can collectively shape the environmental visibility and oversight pressures surrounding supplier operations.

Information processing refers to the gathering, interpreting, and synthesis of information in the context of organizational decision-making (Tushman & Nadler, 1978). Organizational information processing theory (OIPT) suggests that organizations must develop sufficient information-processing capacity to manage uncertainty and complexity effectively (Busse et al., 2017; Galbraith, 1973). In global supply chains, environmental information is frequently fragmented, heterogeneous, and dispersed across organizational and national boundaries, making it difficult for buyers to identify and respond to supplier environmental risks using traditional governance approaches (Wang et al., 2024;

Xie et al., 2026). AI-enabled environmental governance enhances firms' ability to process large volumes of high-dimensional environmental information from multiple sources in real time (Tian et al., 2025). The governance implications of buyer AI-enabled environmental governance may extend beyond focal buyer firms. For example, through AI-enabled systems, buyers can more effectively integrate distributed environmental information, identify potential environmental risks, and enhance environmental visibility across dispersed supply networks. Because suppliers are simultaneously embedded in networks involving multiple buyers, exposure to buyers with stronger AI-enabled environmental information-processing capabilities can collectively intensify environmental oversight at the supplier level. As a result, the likelihood that problematic supplier environmental practices are detected before escalating into public controversies is increased.

In addition to enhancing buyers' environmental information-processing capacity, buyer AI-enabled environmental governance may also influence suppliers through signaling mechanisms. Signaling theory suggests that firms communicate priorities and governance intentions through observable organizational practices and strategic investments under conditions of information asymmetry (Connelly et al., 2011; Connelly et al., 2025; Spence, 1973; Vinayavekhin et al., 2026). Buyers' implementation and disclosure of AI-enabled environmental governance practices—such as AI-based environmental analytics, automated screening systems, and real-time environmental data integration—serve as credible signals of stronger environmental oversight capability. Because these investments are costly and technologically sophisticated, suppliers may perceive heightened scrutiny and a greater likelihood that environmental violations will be detected. These perceptions increase deterrence against environmental misconduct and incentivize suppliers to improve environmental practices.

Taken together, OIPT explains how buyer AI-enabled environmental governance enhances environmental information-processing capacity across dispersed supply networks, whereas signaling theory explains how suppliers interpret buyers' AI-enabled governance investments as signals of heightened governance pressures. Both mechanisms suggest that suppliers exposed to stronger buyer AI-enabled environmental governance are less likely to experience environmental controversies. Therefore, we formally hypothesize:

***H1:*** *Suppliers' exposure to buyers' AI-enabled environmental governance is negatively associated with supplier environmental controversies.*

### 2.3 The Moderating Role of Supplier-Country AI Readiness

The effectiveness of suppliers' exposure to buyers' AI-enabled environmental governance may also depend on the technological environment of supplier countries. Organizational information processing theory (OIPT) suggests that interorganizational information-processing effectiveness depends not only on firms' internal systems but also on the broader environment that facilitates or constrains information acquisition, transmission, and interpretation (Busse et al., 2017). Supplier-country AI readiness reflects the extent to which technological and data infrastructures support the deployment and use of AI-enabled systems (Vann Yaroson et al., 2026), thereby shaping the effectiveness of buyers' AI-enabled environmental information processing.

Specifically, in supplier countries with higher AI readiness, buyers can more effectively deploy AI-enabled environmental governance systems to collect, transmit, process, and analyze supplier-related environmental information. Stronger technological and data infrastructures facilitate real-time communication, data integration, traceability, and cross-boundary coordination (Vann Yaroson et al., 2026), improving the speed and quality of environmental monitoring. As suppliers are exposed to governance systems from multiple buyers, higher supplier-country AI readiness further enhances the interoperability and effectiveness of these mechanisms. In contrast, lower AI readiness may constrain the effectiveness of buyers' AI-enabled information-processing capabilities owing to weaker digital infrastructure and lower technological compatibility.

Therefore, supplier-country AI readiness is expected to strengthen the negative relationship between suppliers' exposure to buyers' AI-enabled environmental governance and supplier environmental controversies. We formally propose:

***H2**: Supplier-country AI readiness positively moderates the negative relationship between suppliers' exposure to buyers' AI-enabled environmental governance and supplier environmental controversies, such that the relationship is stronger in countries with higher levels of AI readiness.*

**2.4 The Moderating Role of Supplier-Country Regulatory Quality**

The effectiveness of suppliers' exposure to buyers' AI-enabled environmental governance may also depend on the regulatory quality of supplier countries. Regulatory quality reflects the ability of governments to formulate and implement effective policies and regulations, as well as the institutional capacity to support compliance and responsible conduct (Agostino et al., 2026). In countries with higher regulatory quality, firms typically face stronger institutional expectations regarding environmental responsibility and greater external scrutiny of corporate conduct.

From a signaling perspective, buyers' AI-enabled environmental governance signals a stronger environmental oversight capability. Suppliers exposed to buyers with strong AI-enabled environmental governance may interpret such signals as indications that environmental misconduct is more likely to become visible and attract stakeholder attention. These signals become more credible and salient in countries with higher regulatory quality, where institutional environments reinforce expectations for environmental compliance and responsible conduct. Stronger regulatory institutions also enhance environmental disclosure quality and data transparency (Zhu & Sarkis, 2007), further facilitating buyers' AI-enabled processing of supplier environmental information.

By contrast, in countries with lower regulatory quality, weaker institutional support for transparency and environmental compliance may reduce the credibility and salience of buyers' AI-enabled environmental governance signals. Suppliers in such environments may perceive lower risks of detection, stakeholder scrutiny, and reputational consequences for environmental misconduct, weakening the deterrent effect of buyer AI-enabled environmental governance. Lower regulatory quality may also limit environmental transparency and reliable information availability, constraining the effectiveness of AI-enabled monitoring across supply chains.

Accordingly, the negative relationship between suppliers' exposure to buyers' AI-enabled environmental governance and supplier environmental controversies should be stronger when supplier-country regulatory quality is high. We formally propose:

***H3:*** *Supplier-country regulatory quality positively moderates the negative relationship between suppliers' exposure to buyers' AI-enabled environmental governance and supplier environmental controversies, such that the relationship is stronger in countries with higher levels of regulatory quality.*

## 3. Methodology

### 3.1 Sample and Data Sources

We focus on foreign suppliers of non-financial, U.S.-listed buyers to ensure consistency in buyer-country market environments and regulatory conditions, thereby enhancing comparability across observations. U.S. firms provide a particularly relevant setting for several reasons. First, U.S. multinational corporations are among the most active participants in global outsourcing and occupy central positions in global supply chains. Second, U.S. firms have been at the forefront of adopting AI-enabled sustainability governance practices, particularly following the rapid diffusion of generative AI technologies after 2022. Third, U.S. buyers often exert substantial monitoring and governance pressure on overseas suppliers through sustainability standards and audits.

Our sample period spans from 2020 to 2024. We begin in 2020 because AI-related corporate disclosures and organizational adoption became increasingly prevalent amid the broader wave of enterprise AI diffusion around this period. In addition, country-level AI readiness indicators exhibited a marked upward shift after 2019, suggesting that firms entered a substantially different technological environment from 2020 onward. The sample ends in 2024 because 2025 SEC filings were not yet fully available at the time of data collection, and including an incomplete filing year could introduce comparability and coverage biases in disclosure-based textual measures.

Our unit of analysis is the supplier-year level because suppliers are frequently connected to multiple buyers simultaneously. Consistent with our governance spillover perspective, we focus on the overall governance influence exerted by buyers' AI-enabled environmental governance rather than isolated dyadic relationships. Accordingly, we aggregate buyer-level AI-enabled environmental governance exposure to the supplier level using buyer-importance weights.

We constructed a novel dataset by integrating multiple sources, including SEC 10-K filings, FactSet Supply Chain Relationships, RepRisk, the Worldwide Governance Indicators (WGI), Oxford Insights, LSEG Asset4, and other firm- and country-level databases. We began by identifying non-financial U.S.-listed firms in Compustat with available SEC 10-K filings and extracting textual disclosures related to AI-enabled environmental governance from the Item 1 (Business) and Item 7 (MD&A) sections. These disclosures capture firms' use of AI technologies in environmental and sustainability governance activities—including environmental analytics, monitoring, compliance support, and environmental risk management—rather than supplier-specific governance systems. Consistent with our spillover argument, the disclosures are intended to capture firms' investments and orientation toward AI-enabled environmental governance capabilities at the organizational level. Prior literature suggests that the Business section reflects firms' strategic orientation and long-term capabilities, serving as an important signal of organizational commitment, whereas the MD&A section provides managerial discussion of risks, uncertainties, and operational responses that reflect information-processing and coordination capabilities (Brown & Tucker, 2011; Li, 2008; Loughran & McDonald, 2011; Ma & Deng, 2026; Ma et al., 2023). These sections are therefore well-suited to capture AI-enabled environmental governance orientation and capabilities, which may support environmental risk management and strengthen governance expectations communicated to foreign suppliers. In robustness checks, we also used the full SEC 10-K filings as text sources.

We then identified foreign suppliers of these non-financial, U.S.-listed firms using the FactSet Supply Chain Relationships database, which captures global inter-firm relationships based on public sources such as SEC filings (10-K reports), investor presentations, and press releases (Dong et al., 2026). We focused on first-tier suppliers because their environmental incidents expose buyers to substantial media scrutiny and because buyers can more directly exert influence on these suppliers (L. Wang et al., 2025). We excluded suppliers incorporated in tax haven jurisdictions to mitigate concerns about opaque ownership structures and limited disclosure transparency. We also included buyers' alternative sustainable supply chain practices as controls using the LSEG Asset4 database (formerly Refinitiv Asset4).

Next, we identified negative environmental incidents for each supplier using RepRisk, which tracks 28 predefined ESG-related incident categories across more than 210,000 companies using a combination of artificial intelligence, machine learning, and human analysis (Kach et al., 2025; L. Wang et al., 2025); we describe this database in detail in Section 3.2.1. We obtained supplier-country AI readiness data from Oxford Insights and supplier-country regulatory quality data from the Worldwide Governance Indicators (WGI) to measure our moderating variables.

We supplemented our dataset with additional firm-level characteristics for both buyers and suppliers from the Directory of Corporate Affiliations (DCA), Compustat, and Bureau van Dijk (BvD) databases. We also incorporated country-level control variables from the World Bank database.

By integrating data from these sources and removing observations with missing values for variables of interest, we constructed an unbalanced panel comprising 5,620 firm-year observations from 2,505 foreign suppliers headquartered in 41 countries between 2020 and 2024. Definitions and detailed measurements of all variables are reported in Table 1.

**Table 1.** Variable definitions and measures

| Variables | Definitions | Measures | Reference | Data Source |
|---|---|---|---|---|
| ***Dependent variable*** | | | | |
| Supplier environmental controversies | The media-reported negative environmental incidents, which represent salient manifestations of supply-chain environmental risks in existing operations and supply chain management literature | The sum of environmental incidents for the focal supplier i in year t | Li and Wu, 2020 | RepRisk |
| ***Independent variable*** | | | | |
| Suppliers' exposure to buyer AI-enabled environmental governance | Buyers' disclosure and deployment of AI in environmental governance to which a supplier is exposed | The aggregated buyer AI-enabled environmental governance to the supplier level using buyer–supplier relationship duration as weights in year t–1 (See details in 3.2.2 section) | Adapted from Wu, 2024 | FactSet supply chain relationship and SEC 10-K |
| ***Moderating variables*** | | | | |
| Supplier-country AI readiness | Supplier-country level technical capacity to integrate AI into the delivery of public services and to support broader economic transformation | The annual Government AIR index in the supplier i's country in year t–1 | Vann Yaroson et al., 2026 | Oxford Insights |
| Supplier-country regulatory quality | The overall effectiveness of the institutional and regulatory environment in suppliers' home countries | The WGI Regulatory Quality index for supplier i's country in year t–1 | Agostino et al., 2026 | Worldwide Governance Indicators (WGI) |
| ***Control variables*** | | | | |
| Supplier size | Supplier-related characteristics | The logarithm of the number of employees in year t–1 | Flammer et al., 2019; Kim et al., 2015, and Dong et al., 2026 | DCA, Compustat, BvD, and FactSet supply chain relationship |
| Supplier age | | The logarithm of the established years in year t–1 | | |
| Supplier net assets | | Net assets in year t–1 | | |
| Supplier leverage | | The ratio of debt to equity in year t–1 | | |
| Suppliers' exposure to non-U.S. customers | | The proportion of non-U.S. foreign buyers relative to the supplier's total number of buyers | | |

**Table 1.** (*Continued*)

| Variables | Definitions | Measures | Reference | Data Source |
|---|---|---|---|---|
| ***Control variables*** | | | | |
| Average buyer environmental performance | Buyer-related characteristics | The average, across all of a supplier's buyers, of environmental pillar score in year t–1 | Flammer et al., 2019; Kim et al., 2015 | LSEG Asset4, DCA, Compustat, and BvD |
| Average buyer controversies | | The average, across all of a supplier's buyers, of ESG controversies in year t–1 | | |
| Traditional supply chain monitoring | | The total, across all of a supplier's buyers, of the adoption of traditional approaches in monitoring suppliers in year t–1 | | |
| Average buyer net assets | | The average, across all of a supplier's buyers, of the logarithm of total assets in year t–1 | | |
| Voice and accountability | Supplier-country-level characteristics | Voice and Accountability index for a supplier's country in year t–1, capturing the transparency and media environment | Dong et al., 2026 | Worldwide Governance Indicators (WGI) |
| GDP (logged) | | The logarithm of GDP for a supplier's country in year t–1 | | World Bank database |

### 3.2 Variables

#### 3.2.1 Dependent Variable

Our dependent variable, supplier environmental controversies, captures media-reported negative environmental incidents—salient manifestations of supply-chain environmental risks recognized in the operations and supply chain management literature. We constructed this measure using RepRisk's comprehensive database, widely validated in prior corporate sustainability research (Kölbel et al., 2017). Previous literature acknowledges that such negative incidents can proxy realized and externally observable ESG-related risks and misconduct to a certain extent (Li & Wu, 2020). The RepRisk database offers several advantages relevant to our research design.

First, RepRisk covers over 150,000 news sources across 30 languages, providing comprehensive media coverage from our sample firms' home and host markets. Second, it employs trained analysts who systematically verify, classify, and document incidents (Wang & Li, 2024), ensuring consistency and reliability. Third, RepRisk categorizes incidents according to standardized criteria aligned with major international ESG frameworks, facilitating cross-firm and cross-country comparability.

Specifically, we operationalized our dependent variable as the sum of environmental incidents for each supplier in a given year. In the robustness checks, we also measured it as a weighted sum of environmental controversies for each supplier, with severity and reach levels serving as weights (coded as 1, 2, or 3 according to RepRisk's classification) (Wang et al., 2024).

#### 3.2.2 Key Independent Variable

Our key independent variable, suppliers' exposure to buyer AI-enabled environmental governance, captures the extent to which a supplier's buyers disclose and deploy AI in environmental governance. We measured buyer AI-enabled environmental governance by extracting substantive meaning from text documents (Item 1 and Item 7 in SEC 10-K filings) through content analysis. Consistent with prior research using corporate disclosures to capture digital and AI-related strategic orientation, we conceptualize AI-enabled governance not as firms' overall technological sophistication, but as the extent to which firms disclose the integration of AI-related tools, analytics, and intelligent systems into governance and operational monitoring processes.

We adopted a bag-of-words (BoW) approach, which quantifies the amount of discussion of a given topic by tabulating the frequency of keywords describing that topic. The bag-of-words can be extended to jointly capture discussions on multiple topics. For example, Wu (2024) measured supply chain risk exposure by tabulating the frequency of co-occurrences of two groups of keywords related to "supply chain" and "risk".

Rather than relying on standalone keyword frequencies, we employed a co-occurrence- and proximity-based textual measure to identify discussions in which AI technologies are applied to environmental governance activities. The intuition underlying our measure is that firms placing greater emphasis on AI-enabled environmental governance are more likely to discuss the use of AI technologies in monitoring, managing, optimizing, and controlling environmental issues within their 10-K disclosures. Such discussions should contain AI-related terms jointly appearing with environmental issue terms and environmental governance/practice terms within the same sentence.

We started by compiling three lexicons: AIWords, capturing AI-related technologies and applications; EnvironmentalWords, capturing environmental issues and objects; and GovernanceWords, capturing environmental governance, monitoring, compliance, and management practices.

The AIWords lexicon was constructed by integrating established AI dictionaries from prior literature (Dahlke et al., 2024; Tian et al., 2025), including terms such as "artificial intelligence," "machine learning," "neural network," and "deep learning". We supplemented the dictionary with generative-AI-era terms (e.g., "generative AI," "large language model") and excluded ambiguous standalone acronyms (e.g., "AR," "BI," "CNN") to reduce false positives. By contrast, constructing the EnvironmentalWords and GovernanceWords lexicons was more challenging because existing dictionaries primarily operationalize corporate environmental engagement and practices broadly or focus on the readability, sentiment, and quality of sustainability and ESG reports (Du & Yu, 2021; Loughran & McDonald, 2011). Thus, we constructed these two lexicons by deriving seed words from prior environmental management literature, expanding them with inflectional and closely related lexical variants, and manually validating all candidate terms against their usage in the 10-K corpus to ensure contextual relevance and construct validity; terms that predominantly generated false-positive matches in non-environmental contexts were excluded. For EnvironmentalWords, we derived words such as "emission," "carbon," "wastewater," "pollutant," "contamination," "energy efficiency," "recycling," "climate risk," and "resource utilization". For GovernanceWords, we derived words such as "monitoring," "auditing," "compliance," "mitigation," "optimization," "oversight," "inspection," "tracking," "process control," and "management system". The three dictionaries are provided in Appendix A.

Prior to textual analysis, we removed non-substantive content (e.g., HTML artifacts, tables, headers, and extremely short sentences) following standard SEC text preprocessing procedures to improve measurement precision. For each document k, we identified all sentences containing at least one AI-related term, one environmental issue term, and one governance/practice term simultaneously. We then computed:

$$AI\ enabled\ environmenatl\ governace_k = \frac{\text{S_k}}{\text{TotalSentences_k}} \qquad (1)$$

where S_k denotes the number of sentences satisfying the triple co-occurrence condition in document k, and TotalSentences_k denotes the total number of sentences in the document. For ease of interpretation, this ratio is multiplied by 1,000 (i.e., qualifying sentences per 1,000 sentences).

This co-occurrence-based design mitigates false positives arising from standalone discussions of AI technologies or environmental topics by requiring semantic proximity among all three dimensions. For example, a sentence such as "the company utilizes machine learning technologies to monitor carbon emissions and optimize wastewater treatment processes" would be identified by our measure because the AI-related term ("machine learning") appears in close proximity to both environmental issue terms ("carbon emissions," "wastewater treatment") and governance/practice terms ("monitor," "optimize"). We manually validated a random subset of firm reports to ensure that identified keywords reflected substantive AI-related governance applications rather than symbolic rhetoric.

Lastly, following prior supply chain research, we aggregated buyer AI-enabled environmental governance to the supplier level using buyer–supplier relationship duration as weights to account for differences in the intensity and persistence of buyer influence. We further considered buyers' net assets as an alternative weighting scheme to capture buyer economic importance.

**3.2.3 Moderating Variables**

Our first moderating variable, supplier-country AI readiness, captures the supplier-country-level technical capacity to integrate AI into public service delivery and to support broader economic transformation, thereby influencing buyers' information-processing capacity under AI-enabled sustainability governance. We measured supplier-country AI readiness using the annual Government AIR index from Oxford Insights (Vann Yaroson et al., 2026). This index constructs an overall AIR_Score for each country annually based on three equally weighted composite pillars: AIR Government, which captures institutional capacity and regulatory preparedness; AIR Technology Sector, which reflects the maturity and density of the domestic technology ecosystem; and AIR Data and Infrastructure, which assesses the quality and accessibility of data systems and digital infrastructure. This measurement is consistent with our theoretical argument that digital and data infrastructure conditions would enhance the information-processing capacity buyers derive from AI-enabled sustainability governance.

Our second moderating variable, supplier-country regulatory quality, captures the overall effectiveness of the institutional and regulatory environment in suppliers' home countries in a given year. We measured this construct using the Regulatory Quality indicator from the Worldwide Governance Indicators (WGI), developed by the World Bank. This indicator reflects perceptions of the government's ability to formulate and implement sound policies and regulations that support private-sector development and effective governance. Higher values indicate stronger regulatory quality and more effective institutional enforcement environments. The WGI Regulatory Quality measure is widely used in international business and institutional research as a country-level proxy for regulatory effectiveness and governance capacity (Agostino et al., 2026).

#### 3.2.4 Control Variables

To minimize potential confounding effects, we selected control variables based on the sustainability and sustainable supply chain management literature (Dong et al., 2026; Flammer et al., 2019; Kim et al., 2015). On the supplier side, we controlled for factors associated with suppliers' resources and environmental management capacity, including supplier size (log number of employees), age (log years since founding), net assets, and leverage (debt-to-equity ratio). Because suppliers may also be influenced by customers from other countries, we additionally controlled for suppliers' exposure to non-U.S. customers, measured as the proportion of non-U.S. foreign buyers relative to the supplier's total number of buyers.

We further controlled for buyer characteristics that may independently shape suppliers' ESG outcomes. Specifically, we included the average environmental score and average ESG controversies of a supplier's buyers, obtained from LSEG Asset4. In addition, we controlled for an alternative governance channel that may influence suppliers' ESG performance, namely traditional supply chain monitoring, whereby buyers monitor suppliers' environmental issues through surveys, audits, supplier site visits, and questionnaires. We measured this variable as the aggregate adoption of traditional supplier monitoring approaches across all of a supplier's buyers. We also controlled for average buyer net assets to account for buyers' resource capacity and influence.

At the country level, we controlled for institutional and environmental conditions in suppliers' home countries. Specifically, we included Voice and Accountability from the Worldwide Governance Indicators (WGI) to capture the transparency and media environment that may affect public scrutiny of environmental misconduct. We further controlled for GDP (logged).

**3.3 Model Specification**

Given that the dependent variable is a nonnegative count variable characterized by substantial right-skewness and a large proportion of zeros, we employed Poisson pseudo-maximum likelihood (PPML). PPML is well-suited for count-type outcomes and remains consistent under general forms of heteroskedasticity without requiring the dependent variable to follow a true Poisson distribution (Correia et al., 2020; Silva & Tenreyro, 2006). In contrast to log-linear OLS models, PPML naturally accommodates zero-valued observations and avoids biases associated with log transformations. As is common in high-dimensional fixed-effects PPML estimation, observations that are perfectly predicted by the fixed effects structure may be dropped during estimation. We included supplier-industry and supplier-country fixed effects to absorb time-invariant supplier-industry and supplier-country heterogeneity, and year fixed effects to capture common macro-level shocks and regulatory trends. To mitigate concerns about reverse causality, all independent, moderating, and control variables were lagged by one year. Additionally, we winsorized all continuous variables at the 1st and 99th percentiles to handle potential outliers. Our baseline model specification was as follows:

$$Y_{i,t} = \beta_0 + \beta_1 \cdot AIEnvGOV_{i,t-1} + \boldsymbol{\gamma}'\mathbf{X}_{i,t-1} + \alpha_s + \mu_c + \lambda_t + \varepsilon_{i,t} \qquad (2)$$

where:

$Y_{i,t}$ denotes the media coverage of environmental incidents of supplier i in year t;

$AIEnvGOV_{i,t-1}$ is the aggregated buyer AI-enabled environmental governance faced by

supplier i in year t−1;

$\mathbf{X}_{i,t-1}$ is a vector of control variables lagged by one year;

$\boldsymbol{\gamma}$ is the conformable coefficient vector;

$\alpha_s, \mu_c, \lambda_t$ represent supplier industry-, supplier country-, and year-fixed effects, respectively;

$\varepsilon_{i,t}$ is the error term.

To test the moderating effects of supplier-country factors, we included the independent variable, each moderator, and their interactions simultaneously in subsequent model specifications. This allowed us to isolate moderating effects from direct influences of moderators. Interaction variables were mean-centered to minimize multicollinearity. Specifically, the moderated models were specified as follows:

$$Y_{i,t} = \beta_0 + \beta_1 \cdot AIEnvGOV_{i,t-1} + \beta_2 \cdot M_{i,t-1} + \beta_3 \cdot \left(AIEnvGOV_{i,t-1} \times M_{i,t-1}\right) + \boldsymbol{\gamma}'\mathbf{X}_{i,t-1} + \alpha_s + \mu_c + \lambda_t + \varepsilon_{i,t} \quad (3)$$

where $M_{i,t-1}$ in denotes supplier-country AI readiness (regulatory quality) in year t−1. All other terms are defined consistently with Equation (2).

# 4. Results

## 4.1 Descriptive Statistics

Tables 2 and 3 present the descriptive statistics and correlation matrix, respectively, for the variables used in the analysis. As shown in Table 2, suppliers' environmental controversies range from 0 to 202, indicating substantial variation across suppliers in the sample. Suppliers' exposure to buyer AI-enabled environmental governance ranges from 0 to 4.66. Overall, the variables exhibit sufficient variation to examine the relationship between suppliers' exposure to buyer AI-enabled environmental governance and suppliers' environmental controversies.

Table 3 reports the pairwise correlations among the variables. The negative correlation between suppliers' exposure to buyer AI-enabled environmental governance and suppliers' environmental controversies provides preliminary support for H1 ($r = -0.042$, $p < 0.01$). In addition, all variance inflation factor (VIF) values are below the commonly recommended threshold of 10, suggesting that multicollinearity is unlikely to be significant.

**Table 2.** Descriptive Statistics

| | Mean | SD | Min | Max | VIF |
|---|---|---|---|---|---|
| Supplier environmental controversies | 0.796 | 5.804 | 0 | 202 | |
| Suppliers' exposure to buyer AI-enabled environmental governance | 0.328 | 0.814 | 0 | 4.66 | 1.22 |
| Supplier-country AI readiness | 71.375 | 8.468 | 40.472 | 82.457 | 4.83 |
| Supplier-country regulatory quality | 70.459 | 11.637 | 44.801 | 88.333 | 9.42 |
| Supplier size | 7.475 | 2.036 | 1.386 | 12.078 | 2.29 |
| Supplier age | 38.723 | 27.906 | 5 | 133 | 1.12 |
| Supplier net assets | 13.584 | 2.054 | 8.484 | 18.643 | 2.17 |
| Supplier leverage | 0.528 | 0.209 | 0.09 | 1.351 | 1.08 |
| Suppliers' exposure to non-U.S. customers | 0.053 | 0.143 | 0 | 0.933 | 1.04 |
| Average buyer environmental performance | 0.633 | 0.218 | 0 | 0.934 | 2.56 |
| Average buyer controversies | 0.605 | 0.35 | 0.014 | 1 | 2.02 |
| Traditional supply chain monitoring | 1.668 | 1.723 | 0 | 9 | 1.26 |
| Average buyer net assets | 10.323 | 1.674 | 5.38 | 12.368 | 3.66 |
| Voice and accountability | 67.976 | 18.662 | 27.914 | 90.369 | 4.58 |
| GDP (logged) | 28.45 | 1.124 | 24.897 | 30.539 | 2.48 |

Note: The mean VIF is 2.52.

### 4.2 Main Results

Table 4 reports coefficient estimates from multi-way fixed-effect models used to test our hypotheses. Specifically, Model 1 includes only control variables. Model 2 introduces the primary independent variable, suppliers' exposure to buyer AI-enabled environmental governance, predicting suppliers' environmental controversies next year. Model 3 incorporates supplier-country AI readiness and its interaction term with suppliers' exposure to buyer AI-enabled environmental governance. Model 4 adds supplier-country regulatory quality and its interaction term.

**Table 3.** Correlation Matrix

| | (1) | (2) | (3) | (4) | (5) | (6) | (7) | (8) | (9) | (10) | (11) | (12) | (13) | (14) | (15) |
|---|---|---|---|---|---|---|---|---|---|---|---|---|---|---|---|
| (1) Supplier environmental controversies | 1.000 | | | | | | | | | | | | | | |
| (2) Suppliers' exposure to buyer AI-enabled environmental governance | -0.041 | 1.000 | | | | | | | | | | | | | |
| (3) Supplier-country AI readiness | 0.009 | -0.024 | 1.000 | | | | | | | | | | | | |
| (4) Supplier-country regulatory quality | 0.050 | -0.110 | 0.708 | 1.000 | | | | | | | | | | | |
| (5) Supplier size | 0.183 | 0.055 | -0.119 | -0.139 | 1.000 | | | | | | | | | | |
| (6) Supplier age | 0.105 | -0.005 | -0.020 | 0.083 | 0.239 | 1.000 | | | | | | | | | |
| (7) Supplier net assets | 0.269 | -0.002 | -0.069 | -0.064 | 0.660 | 0.269 | 1.000 | | | | | | | | |
| (8) Supplier leverage | 0.026 | 0.023 | -0.013 | 0.060 | 0.078 | 0.017 | 0.070 | 1.000 | | | | | | | |
| (9) Suppliers' exposure to non-U.S. customers | 0.015 | -0.034 | -0.015 | 0.066 | 0.037 | 0.021 | 0.052 | 0.066 | 1.000 | | | | | | |
| (10) Average buyer environmental performance | -0.092 | 0.124 | -0.141 | -0.190 | -0.124 | -0.108 | -0.195 | 0.035 | -0.007 | 1.000 | | | | | |
| (11) Average buyer controversies | 0.069 | -0.075 | 0.026 | 0.089 | 0.119 | 0.107 | 0.121 | -0.013 | 0.015 | -0.460 | 1.000 | | | | |
| (12) Traditional supply chain monitoring | -0.014 | 0.327 | -0.083 | -0.081 | -0.011 | -0.055 | -0.037 | 0.043 | -0.009 | 0.272 | -0.070 | 1.000 | | | |
| (13) Average buyer net assets | -0.087 | 0.206 | -0.081 | -0.171 | -0.132 | -0.135 | -0.168 | 0.019 | -0.007 | 0.749 | -0.683 | 0.210 | 1.000 | | |
| (14) Voice and accountability | 0.062 | -0.106 | 0.428 | 0.841 | -0.133 | 0.169 | -0.039 | 0.112 | 0.104 | -0.189 | 0.108 | -0.051 | -0.187 | 1.000 | |
| (15) GDP (logged) | -0.047 | 0.085 | 0.210 | -0.365 | 0.089 | -0.035 | 0.031 | -0.102 | -0.107 | 0.036 | -0.025 | -0.013 | 0.066 | -0.484 | 1.000 |

Notes: N = 5,620. Correlations larger than |0.031| are significant at 5%.

**Table 4.** Results of Multi-Way Fixed-Effect Models (PPML)

| | Model 1 | Model 2 | Model 3 | Model 4 |
|---|---|---|---|---|
| H1: Suppliers' exposure to buyer AI-enabled environmental governance | | -0.282*** | -0.310*** | -0.302*** |
| | | (0.081) | (0.080) | (0.082) |
| Supplier-country AI readiness | | | 0.028 | |
| | | | (0.021) | |
| H2: Suppliers' exposure to buyer AI-enabled environmental governance * Supplier-country AI readiness | | | -0.011* | |
| | | | (0.005) | |
| Supplier-country regulatory quality | | | | 0.143** |
| | | | | (0.051) |
| H3: Suppliers' exposure to buyer AI-enabled environmental governance * Supplier-country regulatory quality | | | | -0.014** |
| | | | | (0.005) |
| Supplier size | 0.033 | 0.052 | 0.048 | 0.047 |
| | (0.085) | (0.085) | (0.084) | (0.083) |
| Supplier age | 0.007* | 0.006* | 0.006 | 0.006* |
| | (0.003) | (0.003) | (0.003) | (0.003) |
| Supplier net assets | 1.075*** | 1.045*** | 1.047*** | 1.050*** |
| | (0.146) | (0.144) | (0.144) | (0.143) |
| Supplier leverage | -0.428 | -0.426 | -0.485 | -0.459 |
| | (0.438) | (0.426) | (0.432) | (0.428) |
| Suppliers' exposure to non-U.S. customers | -1.075* | -1.075* | -1.052* | -1.065* |
| | (0.440) | (0.435) | (0.433) | (0.437) |
| Average buyer environmental performance | -0.039 | -0.164 | -0.096 | -0.035 |
| | (0.481) | (0.481) | (0.466) | (0.484) |
| Average buyer controversies | 0.159 | 0.209 | 0.229 | 0.162 |
| | (0.324) | (0.314) | (0.306) | (0.316) |
| Traditional supply chain monitoring | -0.033 | -0.004 | -0.008 | -0.012 |
| | (0.027) | (0.028) | (0.028) | (0.029) |
| Average buyer net assets | 0.101 | 0.124 | 0.123 | 0.102 |
| | (0.102) | (0.102) | (0.099) | (0.102) |
| Voice and accountability | -0.000 | 0.009 | -0.064 | 0.020 |
| | (0.074) | (0.073) | (0.071) | (0.070) |
| GDP (logged) | -0.770 | -0.786 | -1.088 | -1.011 |
| | (0.890) | (0.874) | (0.654) | (0.820) |
| _cons | 3.539 | 3.431 | 16.800 | 8.928 |
| | (28.359) | (27.622) | (21.177) | (25.994) |
| No. of observations | 5,620 | 5,620 | 5,620 | 5,620 |
| Pseudo R2 | 0.734 | 0.737 | 0.739 | 0.738 |
| Supplier industry-fixed effect | Yes | Yes | Yes | Yes |
| Supplier country-fixed effect | Yes | Yes | Yes | Yes |
| Year-fixed effect | Yes | Yes | Yes | Yes |

Notes: Robust standard errors, clustered at the supplier level, in parentheses. †p < .10; *p < .05; **p < .01; ***p < .001.

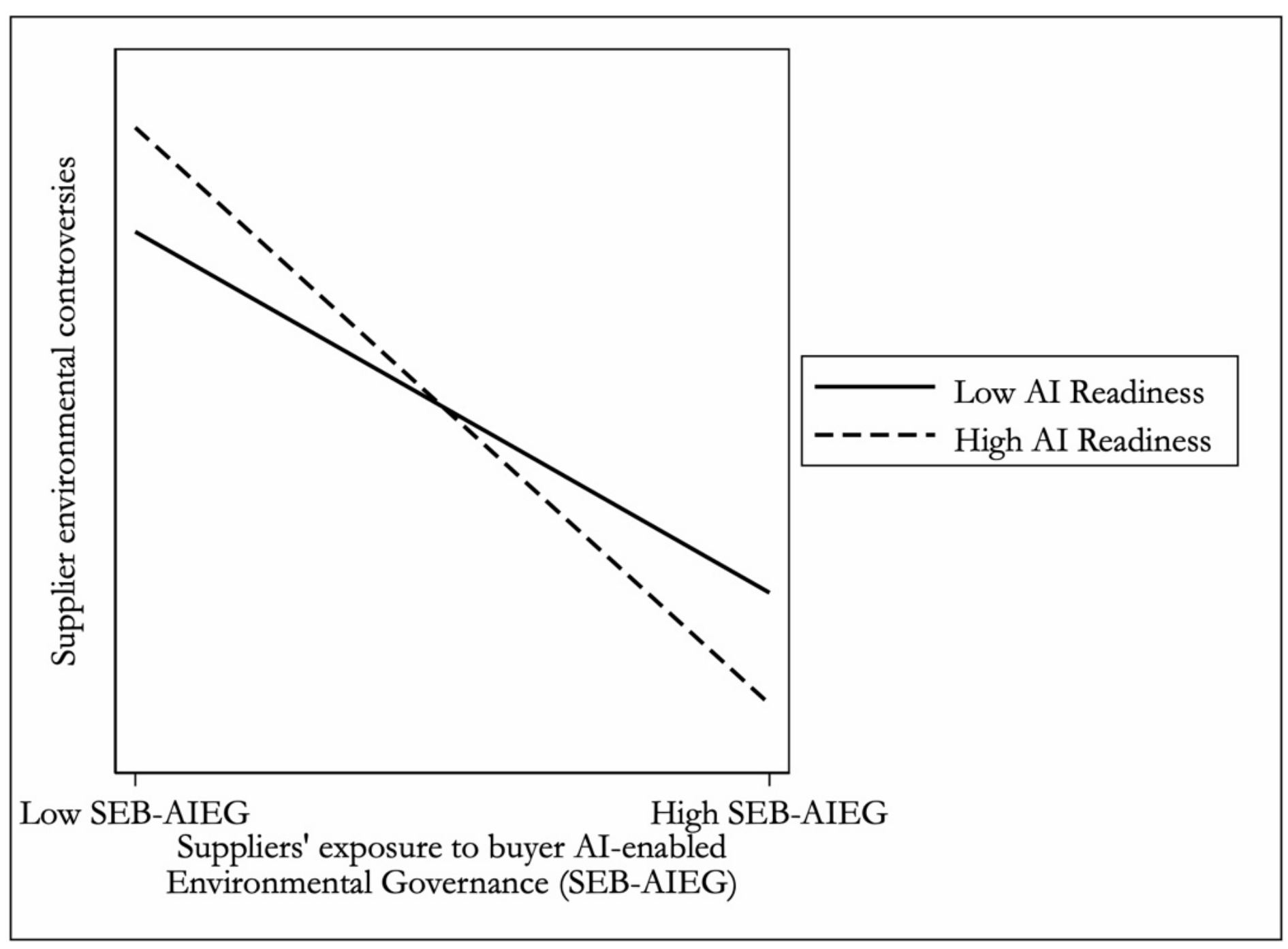

**Figure 2.** The moderating effect of supplier-country AI readiness on the relationship between suppliers' exposure to buyer AI-enabled environmental governance and supplier environmental controversies

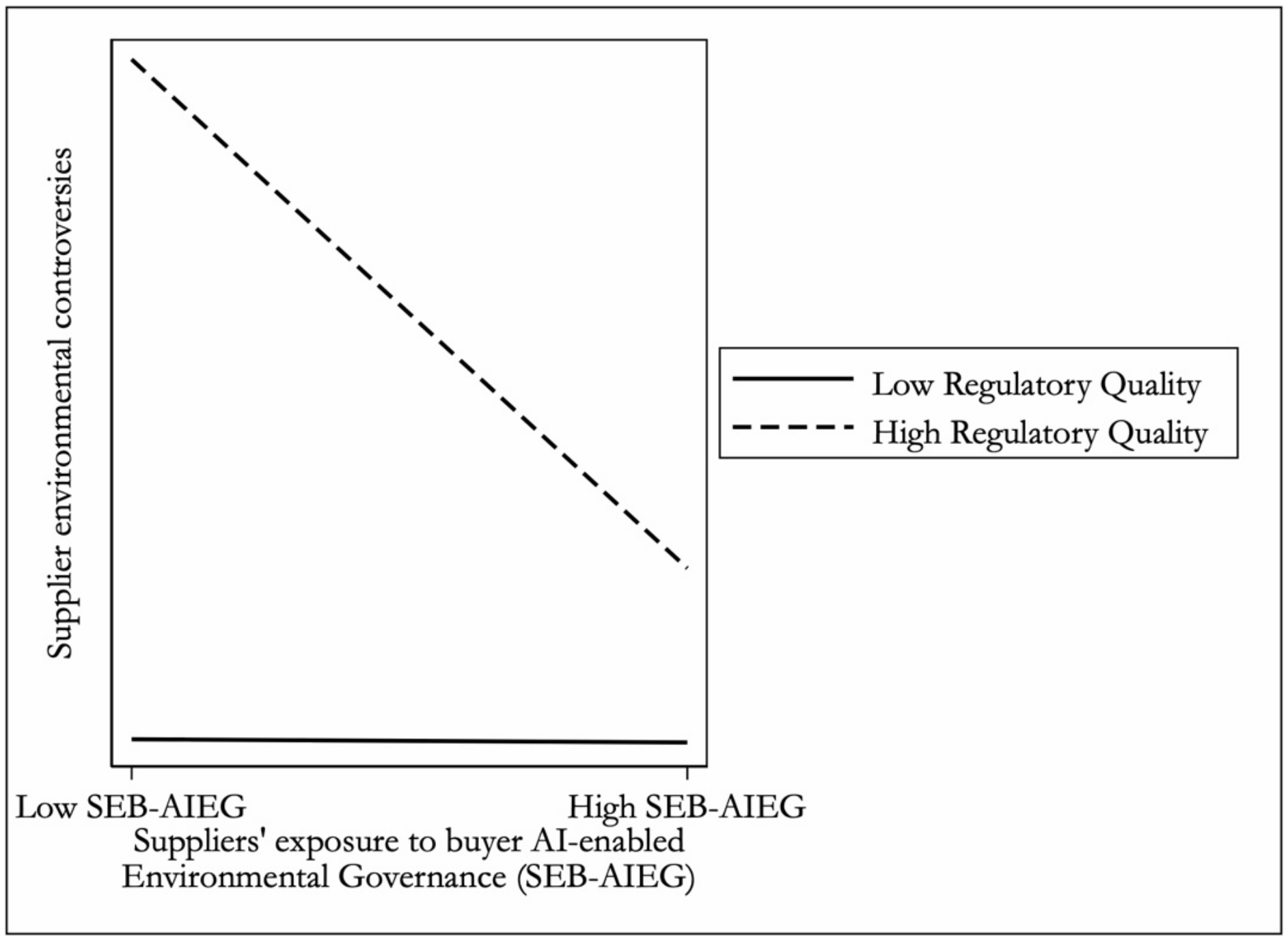

**Figure 3.** The moderating effect of supplier-country regulatory quality on the relationship between suppliers' exposure to buyer AI-enabled environmental governance and supplier environmental controversies

Hypothesis 1 predicts that suppliers' exposure to buyer AI-enabled environmental governance is negatively associated with suppliers' environmental controversies next year. Supporting this prediction, Model 2 shows a significant negative coefficient for suppliers' exposure to buyer AI-enabled environmental governance ($\beta = -0.282$; $p < 0.001$). Thus, Hypothesis 1 is fully supported.

Hypothesis 2 predicted that supplier-country AI readiness enhances the negative relationship between suppliers' exposure to buyer AI-enabled environmental governance and suppliers' environmental controversies next year. In Model 3, the interaction term between suppliers' exposure to buyer AI-enabled environmental governance and supplier-country AI readiness is negatively and statistically significant ($\beta = -0.011$; $p <0.05$), providing support for Hypothesis 2.

Hypothesis 3 predicted that supplier-country regulatory quality enhances the negative relationship between suppliers' exposure to buyer AI-enabled environmental governance and suppliers' environmental controversies next year. Consistent with this expectation, Model 4 indicates that the interaction term suppliers' exposure to buyer AI-enabled environmental governance and supplier-country regulatory quality is significantly negative ($\beta = -0.014$; $p <0.01$), fully supporting Hypothesis 3.

Fig. 2 and Fig. 3 visually illustrate these moderating relationships. Consistent with the negative interaction in Model 3, Fig. 2 shows that the negative relationship between suppliers' exposure to buyer AI-enabled environmental governance and suppliers' environmental controversies next year is stronger when supplier-country AI readiness is high. Similarly, Fig. 3 illustrates that this negative relationship is stronger when supplier-country regulatory quality is high.

**4.3 Robustness Checks**

To verify the robustness of our main findings, we conducted comprehensive supplementary analyses employing alternative measures, samples, and model specifications. These analyses consistently corroborated our primary results (Table 4). Detailed outcomes for each robustness test are provided in Online Appendix B.

#### 4.3.1 Alternative Measures for Key Variables

We employed alternative measures for both the dependent and independent variables to assess the robustness of our findings. First, we remeasured the dependent variable as a weighted sum of supplier environmental controversies, where each controversy was weighted by its severity and reach levels based on RepRisk classifications (coded as 1, 2, or 3). Severity reflects the consequences, scope, and underlying causes of a risk incident, including whether it is accidental, negligent, intentional, or systematic. Reach captures the influence and prominence of the information source, ranging from local media and NGOs to globally recognized media outlets. Because severe incidents are more likely to attract broader media attention, severity and reach are typically highly correlated. This weighted measure captures not only the occurrence but also the intensity and visibility of supplier environmental controversies that may expose buyers to supply-chain environmental risks. We reran all models, and the results were robust (B1).

Second, we reconstructed the independent variable by weighting buyer firms according to their net assets rather than buyer–supplier relationship duration. The results remained consistent with the baseline findings (B2). Furthermore, we constructed an alternative measure of the independent variable based on the full text of firms' SEC 10-K filings. The results remained directionally consistent under this alternative operationalization (B3), although the main effect attenuated to marginal significance, consistent with the full text introducing boilerplate risk-factor language that adds noise to the measure.

#### 4.3.2 Alternative Samples

To ensure that our findings were not driven by sample composition, we conducted several additional analyses using alternative samples. First, we excluded countries with fewer than 10 supplier-year observations to mitigate potential biases arising from sparse country-level representation. Second, we excluded industries with fewer than 100 observations to reduce the influence of thinly represented industries. Third, we restricted the sample to supplier-year observations with positive environmental controversy counts to examine whether the results were sensitive to the large number of zero outcomes in the dependent variable. See B4 to B6 in Appendix B. The results remained largely consistent with the baseline findings. In B6, where the sample is restricted to supplier-years with at least one controversy (662 observations, approximately 12% of the baseline sample), the main effect remained negative and significant, whereas the interaction effects attenuated (the AI readiness interaction became insignificant and the regulatory quality interaction remained significant only at the 10% level). This attenuation is consistent with the sharply reduced sample size and statistical power, and with our theoretical logic: the information-processing and deterrence mechanisms should primarily reduce whether controversies occur at all, rather than the intensity of controversies once they have occurred.

#### 4.3.3 Alternative Estimation Methods

Given that the dependent variable is a nonnegative count variable characterized by overdispersion and a large number of zeros, our baseline analyses employed PPML estimation. To ensure that the findings were not sensitive to the choice of estimation method, we re-estimated the models using negative binomial regression (NB). Because negative binomial models do not readily accommodate high-dimensional fixed effects, we included supplier-industry and year fixed effects in the estimations. The results remained qualitatively consistent with the baseline findings (B7). The consistency of the results across PPML and NB specifications suggests that the findings are not driven by distributional assumptions specific to a particular count-data estimator.

#### 4.3.4 Endogeneity

A potential endogeneity concern is that buyers with stronger AI-enabled environmental governance may systematically maintain supply chains with fewer environmental problems, which could bias the baseline estimates. To provide additional evidence and help alleviate this concern, we employed an instrumental variable approach.

Specifically, we used (1) the average number of AI-related bills passed across buyers' states, obtained from the Stanford AI Index, and (2) the average state-level AI Geographic Exposure (AIGE) across buyers' states developed by Felten et al. (2021). The rationale is that a more active state-level AI regulatory environment and AI occupations may encourage firms to adopt AI-related governance practices, while such state-level legislative activity is unlikely to directly affect environmental controversies at the foreign supplier level except through buyer AI-enabled environmental governance. To alleviate concerns that the instrument may proxy for broader state-level environments, we additionally controlled for state-level economic and environmental characteristics, including the average state-level GDP from the U.S. Bureau of Economic Analysis and renewable energy share from the U.S. Energy Information Administration across buyers' states.

The first-stage results indicate that both instruments are significantly associated with suppliers' exposure to buyer AI-enabled environmental governance. The Kleibergen–Paap rk LM statistic rejects the null of underidentification ($LM = 57.44$, $p < 0.001$), and the Kleibergen–Paap rk Wald F statistic equals 26.09, exceeding the Stock–Yogo critical values, suggesting that weak instrument concerns are unlikely. In addition, the Hansen J test is insignificant ($p = 0.928$), providing support for the validity of the overidentifying restrictions. Consistent with the baseline results, the second-stage estimates show that suppliers' exposure to buyer AI-enabled environmental governance is negatively associated with supplier environmental controversies ($\beta = -2.026$, $p < 0.05$), providing additional support for H1.

**5. Discussion**

Drawing on OIPT and signaling theory, this study reveals a negative association between suppliers'

exposure to buyer AI-enabled environmental governance and supplier environmental controversies. Furthermore, supplier-country AI readiness and regulatory quality enhance this negative association. Our findings yield the following theoretical and managerial implications.

**5.1 Theoretical Contributions**

First, we contribute to the literature on AI adoption and sustainability (Rehman et al., 2026). Prior research primarily examines how firms' AI adoption influences their own ESG outcomes (Dimes et al., 2026; Tian et al., 2025). The dominant view suggests that AI facilitates sustainable growth by improving resource allocation, enhancing sustainability-related knowledge processing, increasing information transparency, and fostering innovation (Rehman et al., 2026; Tian et al., 2025; Xiao & Xiao, 2025).

Drawing on insights from AI applications in supply chain management (Pournader et al., 2021), we extend this literature by theorizing and empirically demonstrating that buyer AI-enabled environmental governance generates governance spillovers across global supply chains, thereby reducing foreign suppliers' environmental controversies. Drawing on organizational information processing theory (OIPT), we argue that buyer AI-enabled environmental governance enhances environmental information-processing capacity across geographically dispersed supplier networks. Complementing this perspective, signaling theory explains how suppliers interpret buyers' AI-enabled governance investments as credible signals of heightened environmental scrutiny and governance expectations. Together, these mechanisms explain why suppliers connected to buyers with stronger AI-enabled environmental governance are less likely to experience environmental controversies. We further identify important institutional contingencies: supplier-country AI readiness strengthens the effectiveness of buyers' information-processing capabilities by enhancing the technical conditions necessary for AI-enabled governance, whereas supplier-country regulatory quality enhances the credibility and enforceability of buyers' governance signals while also facilitating more effective environmental information processing.

Second, we contribute to the emerging literature on AI-enabled sustainability governance (Rehman et al., 2026; Xiao & Xiao, 2025) by developing a firm-level measure of AI-enabled environmental governance using sentence-level textual analysis of SEC 10-K filings. Existing studies have primarily relied on dictionary-based approaches to measure either AI adoption or sustainability-related disclosure separately (Babina et al., 2024; Mittelbach-Hörmanseder et al., 2021; Tian et al., 2025). In contrast, we construct a measure that specifically captures AI-enabled environmental governance by analyzing firms' SEC 10-K filings (Item 1 and Item 7).

Specifically, we identify sentences in which AI-related terms co-occur with environmental issue terms and environmental governance terms, thereby capturing the joint disclosure of AI use and environmental governance rather than either dimension alone. We then aggregate this buyer-level measure to the supplier level through buyer–supplier relationship weights. This design offers a scalable and replicable template for measuring AI-enabled environmental governance in future research on AI and sustainability.

Third, we enrich the sustainable supply chain risk management literature (Damberg et al., 2022; L. Wang et al., 2025) by advancing understanding of how digital governance capabilities reshape environmental risk management in global supply chains. Previous literature has investigated the negative outcomes of sustainable supply chain risks (Jacobs & Singhal, 2017; Mukandwal et al., 2024), and has primarily emphasized conventional governance mechanisms—such as supplier audits, monitoring, supplier termination, and training programs—to mitigate sustainability-related supply chain risks (Hajmohammad & Vachon, 2016; Mateska et al., 2025). While effective, these approaches are often resource-intensive, episodic, and difficult to scale across geographically dispersed supplier networks, which may seek digital solutions (Saberi et al., 2019; Sarkis et al., 2026).

We extend this literature by theorizing AI-enabled environmental governance as a scalable, data-driven governance capability that complements traditional supplier oversight mechanisms. AI-enabled governance allows buyers to continuously process large volumes of environmental information, improve visibility across complex supply networks, and identify potential environmental risks earlier and more systematically than conventional monitoring approaches. Our findings suggest that AI-enabled governance not only strengthens existing supplier oversight practices but also shifts sustainable supply chain risk management toward more continuous, proactive, and network-wide forms of environmental governance. Additionally, our results align with the call for incorporating digital capacity multi-stakeholder partnerships comprising firms and regulators to achieve sustainability goals (Eweje et al., 2021; Sajjad et al., 2020).

### 5.2 Managerial Implications

We also offer implications for managers and policymakers seeking to leverage AI for sustainable development. First, buyer firms should adopt and transparently disclose AI-enabled environmental governance practices to strengthen their capacity to collect, integrate, and continuously monitor environmental information across geographically dispersed supply chains. Rather than relying solely on periodic audits or reactive interventions, firms can use AI-enabled tools to detect abnormal environmental patterns, track compliance risks in real time, and direct oversight toward high-risk suppliers and regions. Public disclosure of these practices may further reinforce suppliers' perceptions that environmental misconduct is more likely to be detected and escalated, thereby strengthening preventive compliance incentives throughout the supply chain. Our findings also suggest that AI-enabled governance should complement rather than replace traditional governance mechanisms such as supplier audits, training programs, and corrective-action systems.

Second, policymakers in supplier countries should strengthen both digital and institutional infrastructures that support effective environmental governance. Investments in AI readiness—including digital and data infrastructures—can enhance global buyers' information processing capacities. At the same time, improving regulatory quality through stronger environmental disclosure standards, enforcement capacity, and monitoring institutions can increase the credibility and effectiveness of buyers' environmental governance signals. These institutional conditions make environmental information more transparent, reliable, and actionable, thereby enhancing the effectiveness of AI-enabled monitoring across global supply chains.

Finally, our findings highlight the importance of cross-border collaboration between multinational buyers, suppliers, and policymakers in building scalable systems for environmental oversight. Because environmental controversies often emerge across fragmented international supply networks, effective governance increasingly depends on shared standards for environmental data, interoperable monitoring systems, and coordinated regulatory support across jurisdictions.

**5.3 Limitations and Future Research**

Our study has several limitations, each of which opens avenues for future research. First, our analysis focuses on suppliers linked to U.S. buyer firms. Because U.S. firms operate within relatively advanced institutional, disclosure, and technological environments, the effectiveness of AI-enabled environmental governance observed in this study may not fully generalize to buyers headquartered in other countries. Future research could examine whether the governance effects of buyer AI capabilities differ across home-country institutional contexts, particularly in emerging economies where AI adoption, disclosure norms, and supply chain governance practices may vary substantially.

Second, although our empirical design incorporates extensive controls and multiple robustness analyses, the study primarily establishes correlational rather than fully causal relationships. In particular, data constraints prevent the inclusion of firm fixed effects in some specifications, raising the possibility that unobserved firm characteristics jointly influence both AI-enabled environmental governance and supplier environmental outcomes. Future research could strengthen causal identification by leveraging natural experiments, regulatory shocks, or exogenous variation in AI adoption and environmental disclosure requirements. Survey-based evidence or in-depth interviews could also be used to directly examine the mediating mechanisms proposed in this study, further clarifying how AI-enabled governance influences supplier environmental behavior over time.

Third, although our study links buyer firms with their suppliers, the analysis does not incorporate transaction-level or relationship-specific data that capture the depth and intensity of interfirm exchanges. As a result, we are unable to distinguish whether AI-enabled environmental governance is more effective in relationships characterized by higher transaction volume or greater supplier dependence. Future research could combine buyer–supplier linkage data with contract- or procurement-level information to identify how AI-enabled governance operates within specific interorganizational relationships. Such research may also clarify whether AI-enabled oversight is more effective in embedded and strategically important supplier relationships than in arm's-length exchanges.

## 6. Conclusion

This study highlights the potential role of buyer firms' AI-enabled environmental governance in reducing supplier environmental controversies in global supply chains. Drawing on organizational information processing and signaling theory, we suggest that AI-enabled governance may strengthen buyers' ability to collect, process, and monitor environmental information across organizational and national boundaries, enhancing oversight of supplier environmental practices. The disclosure of AI-enabled environmental governance may also signal stronger environmental scrutiny and increase suppliers' perceived governance pressures for environmental misconduct. Using panel data on overseas suppliers of U.S.-listed firms from 2020 to 2024 and employing multidimensional fixed-effects models, we find a negative association between buyer AI-enabled environmental governance and subsequent supplier environmental controversies. Our findings further suggest that the effectiveness of buyer AI-enabled environmental governance may depend on the broader technological and institutional context of supplier countries. The stronger association observed in countries with higher AI readiness is consistent with the idea that supportive digital infrastructures may facilitate buyers' environmental information processing capabilities. The stronger effect observed in countries with higher regulatory quality may reflect a context in which governance signals are more credible and environmental oversight and information transparency are more effective. Overall, this study provides evidence that buyer-side AI-enabled governance can generate cross-border governance spillovers and may contribute to more sustainable supply chain management.